\documentclass[twocolumn,10pt]{article}
\usepackage[margin=0.6in]{geometry}
\usepackage{tikz}
\usepackage{pgfplots}
\usepackage{breqn}
\usepackage{lipsum}
\usepackage{soul}
\usepackage{nomencl}

\title{AXIAL FLOW FAN PERFORMANCE IN A FORCED DRAUGHT AIR-COOLED HEAT EXCHANGER FOR A sCO$\mathbf{_2}$ BRAYTON CYCLE}
\author{Francois D. Boshoff, Sybrand J. van der Spuy, Johannes P. Pretorius\\
	Solar Thermal Energy Research Group (STERG), Stellenbosch University, South Africa\\
}
\makeatletter
\newif\if@checkentries
\renewenvironment{nomenclature}[1][]
{\if\relax\detokenize{#1}\relax
	\@checkentriesfalse
	\else
	\settowidth{\@tempdima}{#1\quad}%
	\@checkentriestrue
	\fi
	\def\entry##1##2{%
		\sbox\@tempboxa{##1\quad}%
		\if@checkentries
		\ifdim\wd\@tempboxa>\@tempdima
		\@latex@warning{Entry `\unexpanded{##1}' is wider}%
		\@tempdima=\wd\@tempboxa
		\fi
		\else
		\@tempdima=\wd\@tempboxa
		\fi
		\hangindent\@tempdima
		\noindent\makebox[\@tempdima][l]{##1}\ignorespaces##2\par}%
	\section*{Nomenclature}}
{\par\addvspace{12pt}}
\makeatother
\begin{document}
\maketitle
\begin{abstract}
{\it
An axial flow cooling fan has been designed for use in a concentrated solar power plant. The plant is based on a supercritical carbon dioxide (sCO$_2$) Brayton cycle, and uses a forced draft air-cooled heat exchanger (ACHE) for cooling. The fan performance has been investigated using both computational fluid dynamics (CFD) and scaled fan tests.

This paper presents a CFD model that integrates the fan with the heat exchanger. The objective is to establish a foundation for similar models and to contribute to the development of efficient ACHE units designed for sCO2 power cycles. The finned-tube bundle is simplified, with a Porous Media Model representing the pressure drop through the bundle. Pressure inlet and -outlet boundary conditions are used, meaning the air flow rate is solved based on the fan and tube bundle interaction.

The flow rate predicted by the CFD model is 0.5\% higher than the analytical prediction, and 3.6\% lower than the design value, demonstrating that the assumptions used in the design procedure are reasonable. The plenum height is also found to affect the flow rate, with shorter plenums resulting in higher flow rates and fan efficiencies, and longer plenums resulting in more uniform cooling air flow.

}
\end{abstract}

\begin{nomenclature}[$\alpha_{attack}$]
	\entry{$C$}{Factor}
	\entry{$d$}{Diameter}
	\entry{$G$}{Mass velocity}
	\entry{$H$}{Height}
	\entry{$K$}{Coefficient}
	\entry{$N$}{Number of}
	\entry{$n$}{Thickness}
	\entry{$P$}{Power, Pitch}
	\entry{$p$}{Pressure}
	\entry{$Re$}{Reynolds number}
	\entry{$t$}{Thickness}
	\entry{$V$}{Volume}
	\entry{$v$}{Velocity}
	\entry{$\alpha$}{Correction factor}
	\entry{$\Delta$}{Difference}
	\entry{$\eta$}{Efficiency}
	\entry{$\mu$}{Dynamic viscosity}
	\entry{$\rho$}{Density}
	\entry{$_c$}{Critical}
	\entry{$_{F}$}{Fan}
	\entry{$_{Fc}$}{Fan casing}
	\entry{$_f$}{Fin}
	\entry{$_{he}$}{Heat exchanger}
	\entry{$_r$}{Root, rows}
	\entry{$_{rec}$}{Recovery}
	\entry{$_t$}{Transverse}
	\entry{$_{ts}$}{Total-to-static}
	\entry{$_z$}{Z-direction}
\end{nomenclature}

\section{Introduction}
The use of the supercritical carbon dioxide (sCO$_2$) Brayton cycle in concentrated solar power (CSP) is a relatively new research field, which promises clean and efficient energy production. The sCO$_2$ Brayton cycle could theoretically provide higher thermal efficiencies and compact turbomachinery than the more common steam Rankine cycle \cite{Deshmukh2019a}. The low critical point of sCO$_2$ also makes the cycle particularly well suited to dry cooling, leading to reduced  water consumption for the envisaged plant \cite{Brun2017}. This is important when considering that CSP plants are more likely to be located in arid regions. However, while sCO$_2$ provides many benefits, its complex behaviour also makes the design and operation of such a CSP plant more difficult. 

To achieve the highest thermal efficiency, the sCO$_2$ entering the compressor must be just above its critical temperature and pressure. Near this critical point, the properties of sCO$_2$ are extremely sensitive to changes in temperature and pressure, and are therefore difficult to predict. Furthermore, if either temperature or pressure drops below the critical values, the fluid could separate into two phases. A cooling system for an sCO$_2$ Brayton cycle should therefore be designed carefully to ensure uneven- or over-cooling does not occur. However, while numerous research projects can be found in the literature regarding the modelling and optimization of sCO$_2$ power cycles \cite{Dostal2006, Liu2014} and their dry cooling systems \cite{Moisseytsev2016, Lock2019, Abrahams2022}, very few detailed designs of such dry cooling systems can be found. It is therefore unclear how such a system will behave in reality and what will affect its performance.

\begin{figure}[h!]
	\centering
	\includegraphics[width=.9\linewidth]{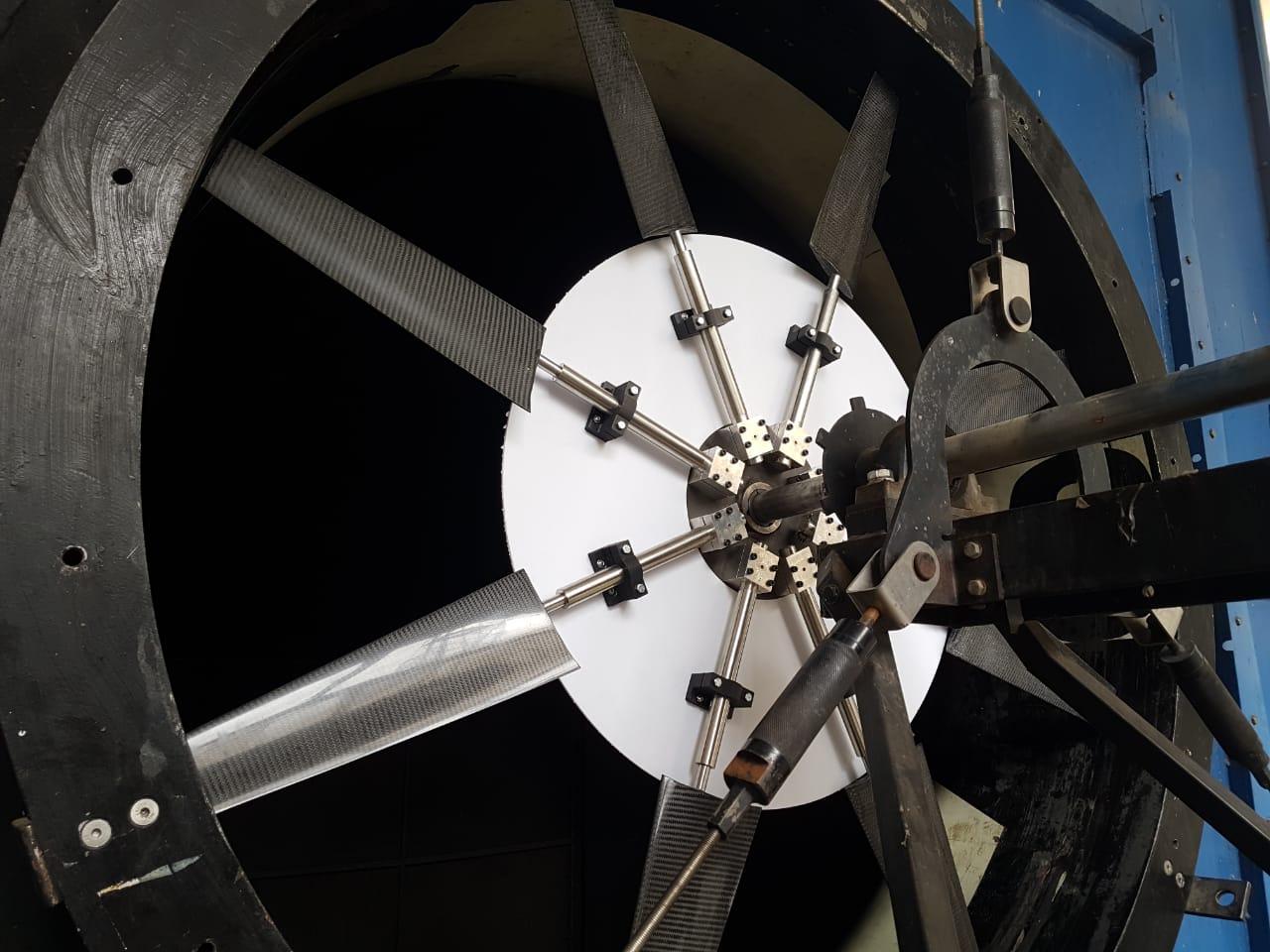}
	\caption{Fan in test facility}
	\label{Figure:FanFromBack}
\end{figure}

To provide an example of such a dry cooling system, Boshoff et al. \cite{Boshoff2022} derived the specifications for a forced draft air-cooled heat exchanger (ACHE) for a 100 MWe sCO$_2$ CSP plant, which uses 6 axial flow fans to provide the necessary cooling air flow. An axial flow cooling fan was then designed specifically for this application, and a scaled model of the fan was manufactured. Due to the unusual duty point of the sCO$_2$ finned-tube bundle, the resulting fan also uses an unusual configuration with a large hub-to-tip ratio of 0.51 and low blade twist, shown in Figure \ref{Figure:FanFromBack}. The performance of the fan was tested extensively using computational fluid dynamics (CFD), with the results being documented in Boshoff et al. \cite{Boshoff2023}. 

\begin{figure}[h!]
	\centering
	\includegraphics[width=.9\linewidth]{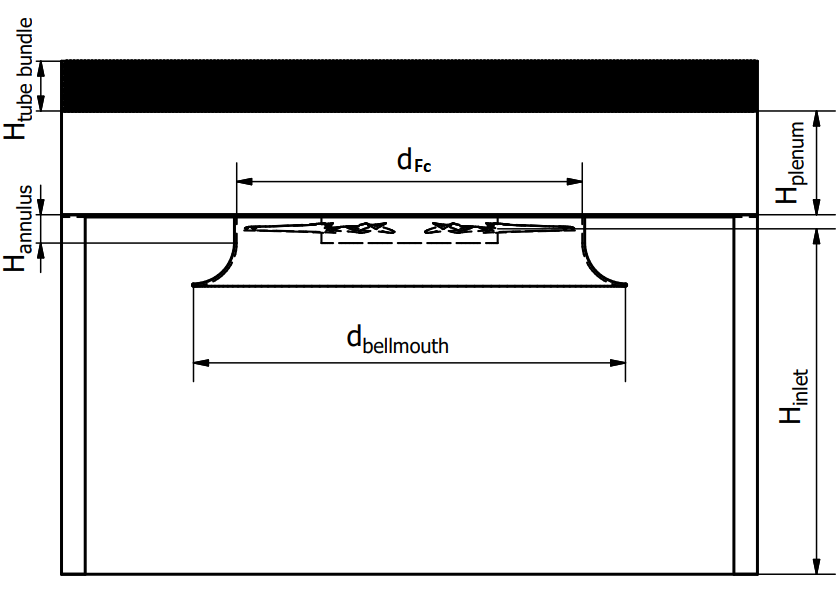}
	\caption{Fan duct layout}
	\label{Figure:ACHELayout}
\end{figure}

This paper continues the previous work \cite{Boshoff2022, Boshoff2023} by investigating the interaction between the fan and air-cooled heat exchanger it was designed for, shown in Figure \ref{Figure:ACHELayout}. Since the construction of a new experimental facility specifically for this purpose is impractical, a CFD-based approach is used instead. Therefore, the aim of the current work is firstly to develop a model of the air-side fan and ACHE interaction with good accuracy, within a practical time frame, and without requiring excessive computational resources. Secondly, the current work aims to provide recommendations for the layout of sCO$_2$-specific ACHEs and cooling fans.

\section{ACHE Layout and Specifications}
While the ACHE consists of six fan ducts, the current work considers only a single fan duct as shown in Figure \ref{Figure:ACHELayout}. The ACHE geometry was mostly specified in the previous work \cite{Boshoff2022}. However, additional dimensions are required for a full three-dimensional model. These include the inlet and plenum chamber heights, which are set to the minimum values recommended by Kr\"{o}ger \cite{Kroger2004}, given in Table~\ref{Table:ACHEgeometry}.

\begin{table}[h!]
	\caption{ACHE specifications}
	\label{Table:ACHEgeometry}
	\centering
	\begin{tabular}{c c}
		& \\
		\hline
		Parameter & Value \\
		\hline
		Air temperature & 15 °C\\
		Air density & 1.209 kg/m$^3$ \\
		Air dynamic viscosity & 1.851$\times$10$^{-5}$ kg/s m\\
		Design air flow rate per fan & 175.7 m$^3$/s\\
		Fan casing diameter & 7.315 m (24 ft)\\
		Fan annulus height & 0.6 m \\
		Bellmouth diameter & 9.144 m ($1.25 \times d_{Fc}$)\\
		Fan inlet height & 7.315 m ($1 \times d_{Fc}$) \\
		Plenum height & 2.195 m ($0.3 \times d_{Fc}$) \\
		Duct width & 14.73 m\\
		Duct length & 10.56 m\\
		Tube bundle height & 1.056 m \\
		Width and depth of pillar & 0.5 m \\
		\hline
	\end{tabular}
\end{table}

\section{Analytical Method}
\subsection{General ACHEs}
\begin{figure}[h!]
	\centering
	\includegraphics[width=.9\linewidth]{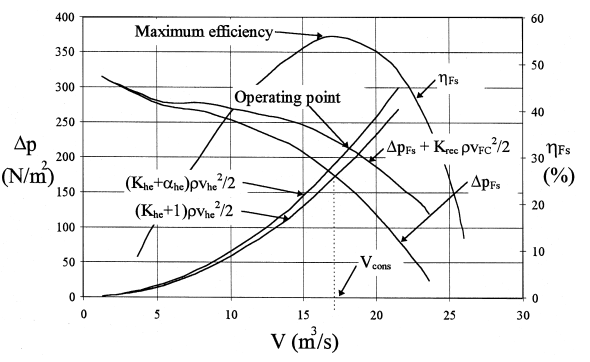}
	\caption{Estimation of operating point, from \cite{Meyer1998}}
	\label{Figure:MeyerKineticRecovery}
\end{figure}

Kr\"{o}ger \cite{Kroger2004} notes that the operating point of a general forced draft ACHE is typically estimated by finding the intersection point between the fan total-to-static pressure rise curve and the system losses curve. These losses include the static pressure drop through the tube bundle and the dynamic pressure lost at the outlet of the bundle. At this point, the useful energy provided to and removed from the air stream are in equilibrium, such that
\begin{equation}
	\Delta p_{ts} = \Delta p_{he} + \frac{\rho v_{he}^2}{2}
	\label{Equation:DraftEquationSimple}
\end{equation}
This approach assumes that the kinetic energy leaving the fan is dissipated in the plenum chamber, and that the cooling air flows uniformly through the tube bundle. Meyer \& Kr\"{o}ger \cite{Meyer1998} conducts experimental tests of such ACHEs, and demonstrates that these assumptions are not always valid, and therefore modifies Equation \ref{Equation:DraftEquationSimple} to account for the fan outlet kinetic energy recovered in the plenum chamber, and the larger kinetic energy lost at the outlet of the tube bundle due to non-uniform flow. The improved equation, known as the draft equation, is given as
\begin{equation}
	\Delta p_{ts} + K_{rec} \frac{\rho v_F^2}{2} = \Delta p_{he} + \alpha_{he} \frac{\rho v_{he}^2}{2}
\end{equation}
where $K_{rec}$ is the kinetic energy recovery factor and $\alpha_{he}$ is the kinetic energy correction factor. These concepts are illustrated in Figure \ref{Figure:MeyerKineticRecovery}. Meyer \& Kr\"{o}ger \cite{Meyer1998} then provides recommendations which can be used to predict $K_{rec}$ and $\alpha_{he}$ in similar heat exchangers, allowing the operating point to be estimated more accurately.

\subsection{Current ACHE and Fan}
Boshoff et al. \cite{Boshoff2022} estimated the static pressure drop of the air flowing through the finned-tube bundle using the empirical correlation by Ganguli \cite{Ganguli1985}, expressed as
\begin{equation}
	\Delta p = \frac{2 G_c^2 N_r}{\rho} \hspace{2 pt} \hspace{2 pt} \left[ 1+\frac{2 e^{-\frac{P_t-d_f}{4 d_r}}}{1+\frac{P_t-d_f}{d_r}} \right] \times ...
\end{equation}
\begin{equation}
	... \times \left[0.021 + 13.6 \frac{d_f - d_r}{Re (P_f - t_f)} + 0.252 \left( \frac{d_f - d_r}{Re (P_f - t_f)} \right)^{0.2}\right]
\end{equation}
However, the parameters of the tube bundle were found to be outside the range of that tested by Meyer \& Kr\"{o}ger \cite{Meyer1998}, and therefore the recommendations regarding the values of $K_{rec}$ and $\alpha_{he}$ could not be used. It was then assumed that no kinetic energy is recovered in the plenum, and the cooling air flows uniformly through the tube bundle. The required fan total-to-static pressure rise was then calculated using the simplified form of the draft equation, as given by Equation \ref{Equation:DraftEquationSimple}.

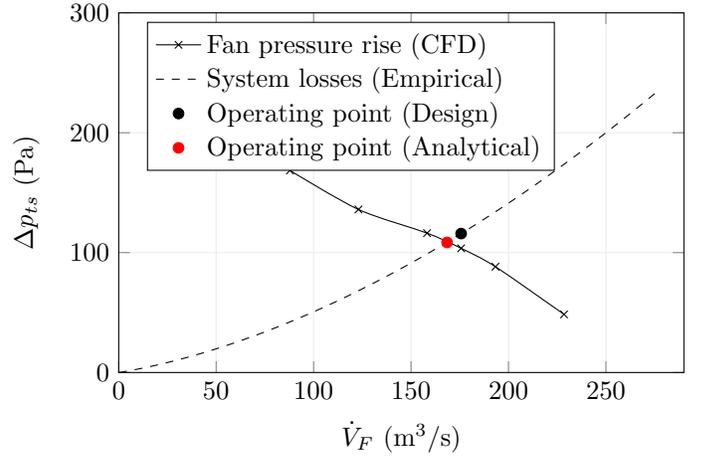
\begin{figure}[h!]
	\centering
	\begin{tikzpicture}
		\begin{axis}[
			width=\linewidth,
			height=.7\linewidth,
			grid=both,
			grid style={draw=gray!15},
			xlabel=$\dot{V}_{F}$ (m$^3$/s),
			ylabel=$\Delta p_{ts}$ (Pa),
			xmin=0,
			xmax=290,
			ymin=0,
			ymax=300,
			legend style={at={(0.05,0.975)},anchor=north west},
			legend cell align = {left}
			]
			\addplot [black,mark=x,smooth]coordinates{(87.87,168.56)(123.0,136.0)(158.2,116.26)(175.7,103.6)(193.3,88.26)(228.5,48.478)};
			\addplot [black,no markers,dashed] table[col sep=comma] {pressure_drop_system.txt};
			\addplot [black,mark=*,only marks]coordinates{(175.7,115.9)};
			\addplot [red,mark=*,only marks]coordinates{(168.5,108.5)};
			\legend{Fan pressure rise (CFD), System losses (Empirical), Operating point (Design), Operating point (Analytical)}
		\end{axis}
	\end{tikzpicture}
	\caption{Design and actual operating points}
	\label{Figure:HeatExchangerPressureDrop}
\end{figure}

After the fan was designed for this operating point and tested, it was found that the total-to-static pressure rise curve is slightly below the design point. Figure~\ref{Figure:HeatExchangerPressureDrop} illustrates the fan total-to-static pressure rise curves obtained from the Test Facility CFD \cite{Boshoff2023}, and the system losses curve, calculated from the empirical correlation \cite{Ganguli1985}. The intersection represents the expected operating point, which is below the design point. The corresponding values are given in Table \ref{Table:TubeBundleFlowRequirements}.

\begin{table}[h!]
	\centering
	\caption{Design and actual flow rate}
	\label{Table:TubeBundleFlowRequirements}
	\begin{tabular}{c c}
		& \\
		\hline
		Parameter & Value\\
		\hline
		Design air flow rate per fan& 175.7 m$^3$/s \\
		Tube bundle average velocity & 1.130 m/s \\
		Tube bundle static pressure drop & 115.1 Pa \\
		Tube bundle outlet dynamic pressure & 0.7713 Pa \\
		Total system losses & 115.9 Pa \\
		Design fan total-to-static pressure rise & 115.9 Pa \\
		Operating point pressure rise & 108.5 Pa \\
		Operating point flow rate & 168.5 m$^3$/s \\
		\hline
	\end{tabular}
\end{table}

\section{Numerical Method}
The meshing, flow modelling and solution methods used in the current work are based on those of \cite{Boshoff2023}, since these showed good correlation to experimental fan tests. The methods are described below, and are implemented in ANSYS Fluent. A sensitivity analysis is provided in Appendix~\ref{Appendix:SensitivityAnalysis}, while alternative methods of modelling the fan are demonstrated in Appendix~\ref{Appending:FanModellingApproach}.

\subsection{Computational Domain and Mesh Generation}
\begin{figure}[h!]
	\centering
	\includegraphics[width=.5\linewidth]{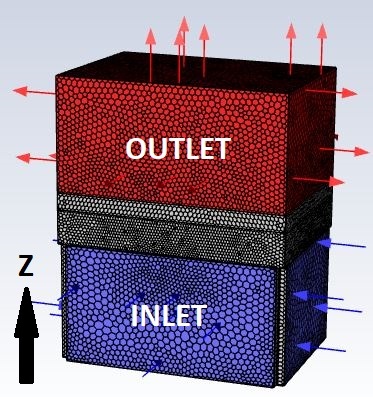}
	\caption{Fan duct mesh}
	\label{Figure:ACHEDomainAssembled}
\end{figure}

The computational domain models a single fan duct as described in Figure \ref{Figure:ACHELayout} and Table \ref{Table:ACHEgeometry}. This domain is divided into five sub-domains: an inlet domain modelling the flow of air past the floor and support pillars, the full fan rotor geometry, the plenum chamber, the tube bundle, and an outlet domain to model the flow leaving the heat exchanger towards the atmosphere. A hexahedral mesh is generated for the fan rotor sub-domain using ANSYS Turbogrid, while tetrahedral meshes are generated for all other sub-domains using ANSYS Meshing. The tetrahedral meshes are then converted to polyhedral meshes in ANSYS Fluent. The assembled mesh is shown in Figure \ref{Figure:ACHEDomainAssembled}.

\subsection{Boundary and Cell Zone Conditions}
All sub-domains are connected using standard mesh interfaces. The atmosphere upstream and downstream of the heat exchanger is modelled by specifiying the inlet and outlet surfaces as pressure inlet and pressure outlet boundary conditions, with the gauge total pressure at the inlet being zero, and the gauge static pressure at the outlet being zero. All surfaces are specified as no-slip wall boundary conditions, with the fan blade and hub surfaces rotating at the fan speed of 147.4 rpm. Finally, the fan rotor sub-domain uses a rotating reference frame which rotates at the fan speed.

\subsection{Flow and Turbulence Modelling}
The Reynolds-averaged Navier-Stokes (RANS) equations are solved, with flow being considered incompressible and steady in the relative frame of reference. Turbulence is modelled using the realizable k-epsilon model and enhanced wall functions.

\subsection{Solution Methods}
ANSYS Fluent's pressure-based, steady flow solver is used with the relative velocity formulation, with the solution methods given in Table \ref{Table:SolutionMethods}. The default under-relaxation factors suggested by ANSYS Fluent are used.

\begin{table}[h!]
	\centering
	\caption{Solution methods and controls used}
	\label{Table:SolutionMethods}
	\begin{tabular}{c c}
		& \\
		\hline
		Parameter & Value\\
		\hline
		P-V coupling scheme & SIMPLE \\
		Gradient & Least Squares Cell Based\\
		Pressure & PRESTO!\\
		Momentum & QUICK\\
		Turbulent kinetic energy & QUICK\\
		Turbulent dissipation rate & QUICK\\
		\hline
	\end{tabular}
\end{table}

\subsection{Tube Bundle Modelling}
Owen \cite{Owen2010} and Van der Spuy \cite{VanderSpuy2011} demonstrate the use of CFD methods for simulating ACHEs. Both models simplify the tube bundle by using a Porous Media Model \cite{ANSYSFluent2013}, which is also used in the current work. 

Therefore, the tube bundle sub-domain is specified as a Porous Zone, which requires specification of an inertial and viscous resistance factor. To calculate these values for the current tube bundle, the pressure drop must firstly be expressed as a second order polynomial. Ganguli's correlation is used to find the pressure drop at air velocities of 80\% and 120\% of the design point flow rate, through which a second order polynomial curve fit is found using MATLAB. The pressure drop is then expressed as
\begin{equation}
	\Delta p_{he} = 46.54 v_{he}^2 + 49.19 v_{he}
\end{equation}
The inertial and viscous resistance factors for the $z$-direction (see Figure \ref{Figure:ACHEDomainAssembled}) are then calculated from the coefficients of this polynomial fit, using
\begin{equation}
	C_z = \frac{2 a}{\rho \Delta n} = 72.92 \hspace{2 pt} \mathrm{{m^{-1}}} \hspace{15 pt} \frac{1}{\alpha_z} = \frac{b}{\mu \Delta n} = 2.517 \times 10^6 \hspace{2 pt} \mathrm{m^{-2}}
\end{equation}
where $\Delta n$ here refers to the thickness of the tube bundle. For the $x$- and $y$- directions, the viscous resistance factors are set to zero, and the inertial resistance factors to a large number, to model the flow-straightening effects of the tube bundle \cite{Owen2010, VanderSpuy2011}.

To test the accuracy of the Porous Media Model, a simple computational domain containing only the tube bundle is used to estimate the static pressure drop at specific velocities. A velocity inlet boundary condition is used to set the velocity through the bundle to 80\%, 100\% and 120\% of the design flow rate (0.9037~m/s, 1.130 m/s and 1.356 m/s), while a pressure outlet boundary condition is used to model the outlet atmosphere. The walls are specified as no-shear. The resulting static pressure drops are compared to that of Ganguli's correlation and the polynomial fit in Table \ref{Table:PorousZoneValidation}, which shows good agreement near the flow rates expected in the current work.
\begin{table}[h!]
	\centering
	\caption{Pressure drop through tube bundle}
	\label{Table:PorousZoneValidation}
	\begin{tabular}{c c c c c}
		& & & \\
		\hline
		Method & 0.9037 m/s & 1.130 m/s & 1.356 m/s \\
	
		\hline
		Ganguli & 82.47 Pa & 115.1 Pa & 152.2 Pa \\
		Polynomial fit & 82.47 Pa & 115.0 Pa & 152.2 Pa \\
		Porous zone & 82.48 Pa & 115.0 Pa & 152.2 Pa \\
		\hline
	\end{tabular}	
\end{table}

\subsection{Fan Modelling}
Owen \cite{Owen2010} simplifies the fan by using a Pressure Jump Model, while Van der Spuy \cite{VanderSpuy2011} simplifies the fan using an Actuator Disk Model. For the current work, the uniformity of the air leaving the fan is under investigation. Therefore, it was decided that the blade, hub and casing geometry should be modelled fully, similar to that of the fan performance simulations in Boshoff et al. \cite{Boshoff2023}. This method allows flow around the large hub, blade pressure pulses, and tip clearance leakage vortexes to be modelled.

The previous fan performance simulations from Boshoff et~al.~\cite{Boshoff2023} used cylindrical inlet- and outlet sub domains, such that the flow around each fan blade would be similar. It was therefore possible to model only a single fan blade with periodic boundary conditions and a rotating frame of reference, where the flow field was considered to be steady. This approach is known as the Frozen Rotor approach. However, for the current computational domain, which features rectangular sub-domains upstream and downstream of the fan, the flow field can no longer be considered periodic or steady. The full fan geometry is therefore modelled for this domain. 

While flow cannot be considered steady in this domain, the solution of a transient model of the flow containing all necessary details was considered to take impractically long. The flow field was therefore assumed to be steady, with the rotor sub-domain being specified as rotating with the fan. The sensitivity of the results to this steady flow assumption is demonstrated by comparing to alternative fan modelling approaches, in Appendix \ref{Appending:FanModellingApproach}.

\subsection{Obtaining Fan Performance}
The cooling air flow rate through the fan duct is obtained by taking the volumetric flow rate through the inlet boundary condition surface. The fan total-to-static pressure rise is obtained by taking the difference between the mass-averaged total pressure at the inlet (always zero, due to the boundary condition) and the mass-averaged static pressure at the interface between the plenum chamber and tube bundle. The fan power consumption is computed by multiplying the torque on the rotating fan blade and hub surfaces with the rotational speed. Finally, the fan total-to-static efficiency is calculated using
\begin{equation}
	\eta_{ts} = \frac{\Delta p_{ts} \dot{V}_F}{P_F}
\end{equation}

\subsection{Convergence Criteria}
\begin{figure}[h!]
	\centering
	\begin{tikzpicture}
		\begin{axis}[
			width=\linewidth,
			height=.6\linewidth,
			xlabel=Iterations ($\times$1000),
			ylabel=Flow rate (m$^3$/s),
			ymin=165,
			ymax=185,
			xmin=0,
			xmax=40,
			grid=both,
			grid style={draw=gray!15},
			legend style={at={(0.95,0.95)},anchor=north east},
			legend cell align = {left}
			]
			\addplot [color = blue,thick,solid] table[col sep=comma,x expr=\thisrowno{0}, y expr=\thisrowno{1}]{ConvergenceNoTipGap.txt};
			\addplot [color = red,thick,solid] table[col sep=comma,x expr=\thisrowno{0}, y expr=\thisrowno{1}]{ConvergenceDesignTipGap.txt};
			\legend{No tip gap, Design tip gap}
		\end{axis}
	\end{tikzpicture}
	\caption{Flow rate convergence}
	\label{Figure:Convergence}
\end{figure}
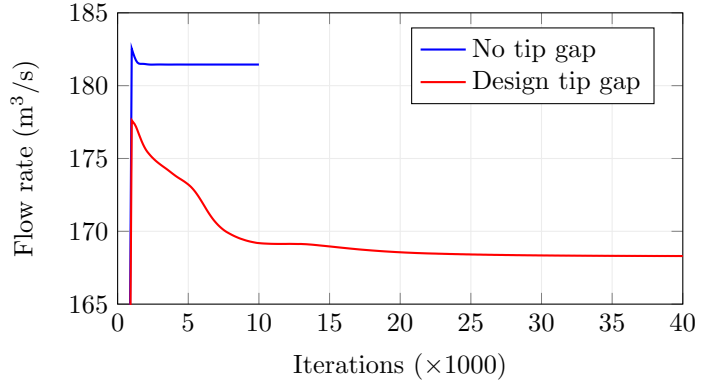

The convergence of the cooling air flow rate of a typical solution process, for both the design tip gap (29.26 mm) and no tip gap cases, is shown in Figure \ref{Figure:Convergence}, while the residuals obtained are given in Table \ref{Table:Convergence}. This comparison demonstrates that the inclusion of the tip gap drastically increases the number of iterations required for convergence. However, to accurately model the operating point and uniformity of the cooling air flow, the tip gap must be modelled. The solution was therefore considered to be completed once the cooling air flow rate has converged to four significant digits. This point was usually obtained after 40 thousand iterations. Due to this long solution process and the large number of simulations required to validate the model, the simulations were completed using the Department of Science and Innovation's Center for High Performance Computing Lengau cluster.

\begin{table}[h!]
	\centering
	\caption{Residuals at completion}
	\label{Table:Convergence}
	\begin{tabular}{c c c}
		& & \\
		\hline
		Parameter & No tip gap & Design tip gap \\
		\hline
		Continuity & 4.180 $\times$10$^{-5}$ & 1.121 $\times$10$^{-4}$ \\
		X-Velocity & 1.298 $\times$10$^{-9}$ & 6.201 $\times$10$^{-8}$ \\
		Y-Velocity & 1.278 $\times$10$^{-9}$ & 6.238 $\times$10$^{-8}$ \\
		Z-Velocity & 1.184 $\times$10$^{-9}$ & 3.116 $\times$10$^{-8}$ \\
		k & 2.667 $\times$10$^{-5}$ & 4.620 $\times$10$^{-5}$ \\
		Epsilon & 8.245 $\times$10$^{-5}$ & 1.199 $\times$10$^{-3}$ \\
		\hline
	\end{tabular}	
\end{table}

\section{Results - Baseline Case}
The results are firstly presented for a baseline case, which uses the geometry from Table \ref{Table:ACHEgeometry}. To ensure the results are reasonable and realistic, this case is compared to the results from the fan test facility simulations done previously \cite{Boshoff2023}.

\subsection{Operating Point}
Table \ref{Table:OperatingPoint} compares the design point, the operating point predicted analytically, and the operating point predicted with the baseline case CFD model. The good agreement between the analytical and CFD results suggest that the assumptions made during the fan design procedure, which were that the kinetic energy recovery and the non-uniformity of the flow leaving the heat exchanger can both be neglected, were reasonable. The reduced cooling air flow rate obtained is therefore considered to be only due to the fan pressure rise curve passing below the design value.

\begin{table}[h!]
	\centering
	\caption{Operating point results}
	\label{Table:OperatingPoint}
	\begin{tabular}{c c c}
		& \\
		\hline
		Method & Flow rate \\
		\hline
		Design point & 175.7 m$^3$/s \\
		Analytical Prediction & 168.5 m$^3$/s \\
		Baseline CFD & 169.4 m$^3$/s \\
		\hline
	\end{tabular}	
\end{table}

\subsection{Pressure Distribution}
\begin{figure}[h!]
	\centering
	\begin{tikzpicture}
		\begin{axis}[
			width=\linewidth,
			height=.5\linewidth,
			ylabel=Pressure (Pa),
			xmin=-5,
			xmax=7,
			ymin=-50,
			ymax=175,
			grid=both,
			grid style={draw=gray!15},
			legend style={at={(0.025,0.975)},anchor=north west},
			legend cell align = {left}
			]
			\addplot [color = black,thick,solid] table[col sep=comma,x expr=\thisrowno{0}, y expr=\thisrowno{1}]{pressure_axial_ACHEPorousFine.txt};
			\addplot [color = red,thick,solid] table[col sep=comma,x expr=\thisrowno{0}, y expr=\thisrowno{3}]{pressure_axial_ACHEPorousFine.txt};
			\addplot [color=blue,thick,solid] table[col sep=comma,x expr=\thisrowno{0}, y expr=\thisrowno{2}]{pressure_axial_ACHEPorousFine.txt};
			\addplot +[color=black,dashed,mark=none] coordinates {(-0.3, -45)(-0.3, 200)};
			\addplot +[color=black,dashed,mark=none] coordinates {(0.3, -45)(0.3, 200)};
			\addplot +[color=black,dashed,mark=none] coordinates {(2.49456, -45)(2.49456, 200)};
			\addplot +[color=black,dashed,mark=none] coordinates {(3.55056, -45)(3.55056, 200)};
			\draw (rel axis cs:0.2,0.15) node[below,align=left] {INLET};
			\draw (rel axis cs:0.4,0.95) node[below,align=left] {FAN};
			\draw (rel axis cs:0.55,0.56) node[below,align=left] {PLENUM};
			\draw (rel axis cs:0.68,0.95) node[below,align=left] {BUNDLE};
			\draw (rel axis cs:0.85,0.15) node[below,align=left] {OUTLET};
			\legend{Static, Dynamic, Total}
		\end{axis}
	\end{tikzpicture}
	\centering
	\begin{tikzpicture}
		\begin{axis}[
			width=\linewidth,
			height=.5\linewidth,
			xlabel=Height (m),
			xmin=-5,
			xmax=7,
			ymin=-50,
			ymax=175,
			grid=both,
			grid style={draw=gray!15},
			ylabel=Pressure (Pa),
			legend style={at={(0.025,0.975)},anchor=north west},
			legend cell align = {left}
			]
			\addplot [color = black,solid,thick] table[col sep=comma,x expr=\thisrowno{0}, y expr=\thisrowno{1}+103.6]{PressureAxialDistribution_Far_TypeADesignTipGap.txt};
			\addplot [color=red,solid,thick] table[col sep=comma,x expr=\thisrowno{0}, y expr=\thisrowno{3}]{PressureAxialDistribution_Far_TypeADesignTipGap.txt};
			\addplot [color = blue,solid,thick] table[col sep=comma,x expr=\thisrowno{0}, y expr=\thisrowno{2}+103.6]{PressureAxialDistribution_Far_TypeADesignTipGap.txt};
			\addplot +[color=black,dashed,mark=none] coordinates {(-0.3, -45)(-0.3, 200)};
			\addplot +[color=black,dashed,mark=none] coordinates {(0.3, -45)(0.3, 200)};
			\draw (rel axis cs:0.2,0.15) node[below,align=left] {SETT. CHAMBER};
			\draw (rel axis cs:0.4,0.95) node[below,align=left] {FAN};
			\draw (rel axis cs:0.65,0.15) node[below,align=left] {OUTLET};
			\legend{Static, Dynamic, Total}
		\end{axis}
	\end{tikzpicture}
	\caption{Pressure distributions through ACHE (top) and test facility (bottom)}
	\label{Figure:BaselinePressureDistribution}
\end{figure}
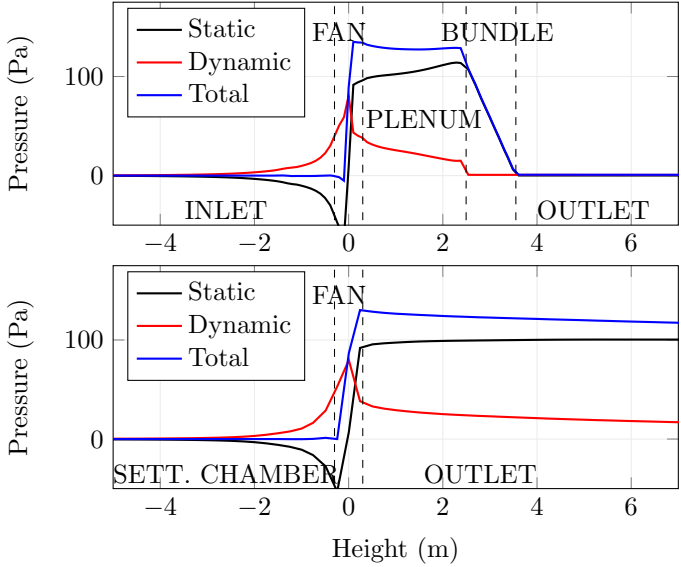

Figure \ref{Figure:BaselinePressureDistribution} (top) illustrates the change in static, dynamic and total pressure of the air as it flows upwards through the heat exchanger, by taking the mass-averaged pressure values for cross-sections at various heights. The resulting pressure distributions are similar to that seen previously in the fan test facility CFD, shown in Figure \ref{Figure:BaselinePressureDistribution} (bottom), where flow enters from a settling chamber and leaves towards the atmosphere. As expected, the pressure is also seen to drop as air flows through the tube bundle.

\begin{figure}[h!]
	\centering
	\includegraphics[width=\linewidth]{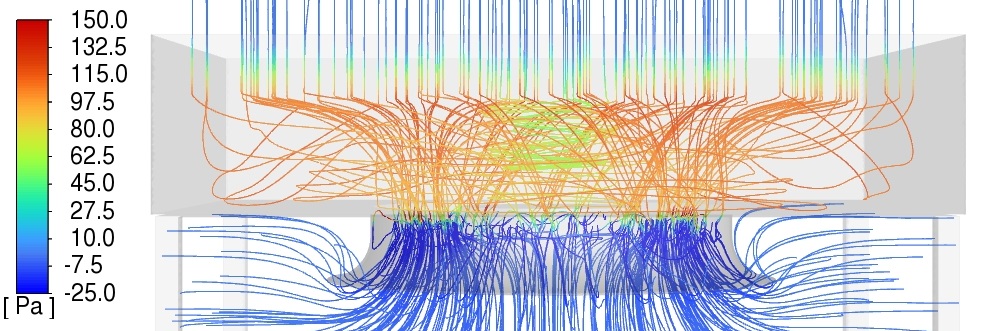}
	\caption{Pathlines colored by static pressure}
	\label{Figure:ACHEPathlines}
\end{figure}

However, an increase in static and total pressure is seen just upstream of the tube bundle. While the static pressure increase could be due to kinetic energy recovery (such that energy is converted from one form to another), the total pressure increase is unusual, since no energy sources should be present at this location. To explain this behavior, Figure \ref{Figure:ACHEPathlines} illustrates the flow field in the plenum chamber with pathlines colored by static pressure. As shown, the pressure of the flow leaving the fan increases as it stagnates upstream of the tube bundle, which causes flow to be deflected outwards and backwards. Since the returned flow has already dissipated some of its energy, it would reduce the average total pressure upstream of the bundle, resulting in what appears to be a downstream increase in total pressure.

\subsection{Fan Performance}
\begin{figure}[h!]
	\centering
	\begin{tikzpicture}
		\begin{axis}[
			width=\linewidth,
			height=0.7\linewidth,
			grid=both,
			grid style={draw=gray!15},
			ylabel=$\Delta p_{ts}$ (Pa),
			xmin=50,
			xmax=250,
			ymin=0,
			ymax=300,
			legend style={at={(0.05,0.975)},anchor=north west},
			legend cell align = {left}
			]
			\addplot [black,mark = x,smooth]coordinates{(87.87,168.56)(123.0,136.0)(158.2,116.26)(175.7,103.6)(193.3,88.26)(228.5,48.478)};
			\addplot [black,no markers,dashed] table[col sep=comma] {pressure_drop_system.txt};
			\addplot [red,mark=*,only marks]coordinates{(169.4,110.5)};
			\legend{Fan pressure rise (Test Facility CFD), System losses (Empirical), ACHE CFD, ACHE CFD Variations}
		\end{axis}
	\end{tikzpicture}
	\centering
	\begin{tikzpicture}
		\begin{axis}[
			width=\linewidth,
			height=0.4\linewidth,
			grid=both,
			grid style={draw=gray!15},
			ylabel=$P_F$ (kW),
			xmin=50,
			xmax=250,
			ymin=20,
			ymax=40,
			legend style={at={(0.4,0.5)},anchor=north west},
			legend cell align = {left}
			]
			\addplot [red,mark=*,only marks]coordinates{(169.4,31.91)};
			\addplot [black,mark =x,smooth]coordinates{(87.87,30.982)(123.0,32.903)(158.2,33.085)(175.7,32.280)(193.3,31.243)(228.5,27.714)};
		\end{axis}
	\end{tikzpicture}
	\centering
	\begin{tikzpicture}
		\begin{axis}[
			width=\linewidth,
			height=0.4\linewidth,
			grid=both,
			grid style={draw=gray!15},
			xlabel=$\dot{V}_F$ (m$^3$/s),
			ylabel=$\eta_{ts}$ (\%),
			xmin=50,
			xmax=250,
			ymin=35,
			ymax=65,
			legend style={at={(0.4,0.5)},anchor=north west},
			legend cell align = {left}
			]
			\addplot [red,mark=*,only marks]coordinates{(169.4,58.66)};
			\addplot [black,mark=x,smooth]coordinates{(87.87,47.804)(123.0,50.84)(158.2,55.58)(175.7,56.39)(193.3,54.61)(228.5,39.97)};
		\end{axis}
	\end{tikzpicture}
	\caption{Comparison of fan performance in ACHE and test facility}
	\label{Figure:HeatExchangerEfficiencyComp}
\end{figure}
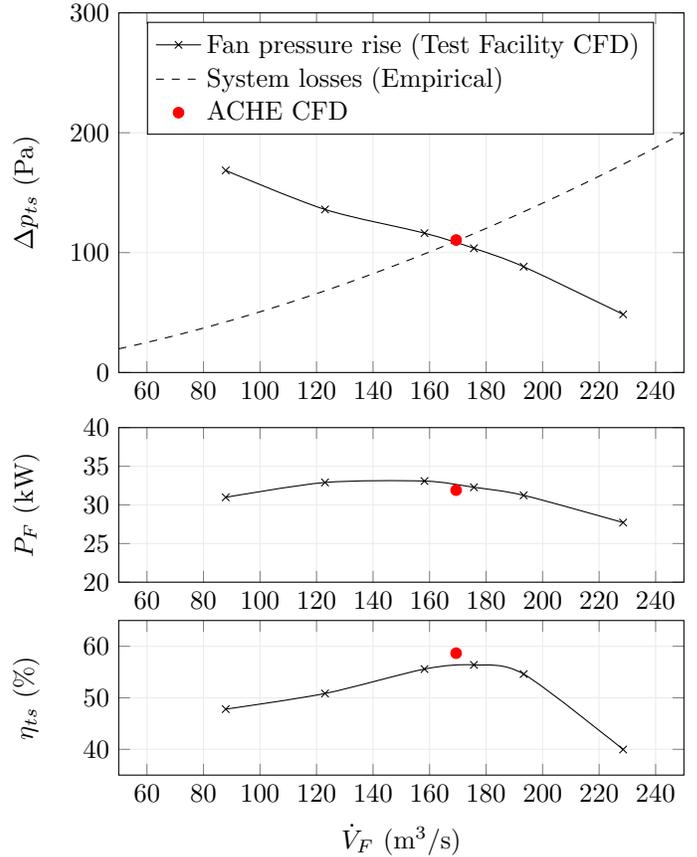

The operating point fan performance is given in Figure \ref{Figure:HeatExchangerEfficiencyComp}, demonstrating that the fan performance is similar to the fan test facility CFD results. The pressure and efficiency is however seen to be slightly higher, while the power consumptions is seen to be slightly lower, possibly due to a small amount of kinetic energy being recoverd in the plenum chamber.

\subsection{Cooling Air Uniformity}
Based on the good agreement obtained between the operating points predicted by the analytical and CFD methods, the assumption of uniform flow through the tube bundle seems reasonable. However, as mentioned previously, the non-uniformity of the cooling air could have a significant effect on the overall behavior of the heat exchanger, since it could lead to the sCO$_2$ in certain tubes being over- or under-cooled.

\begin{figure}[h!]
	\centering
	\includegraphics[width=.95\linewidth]{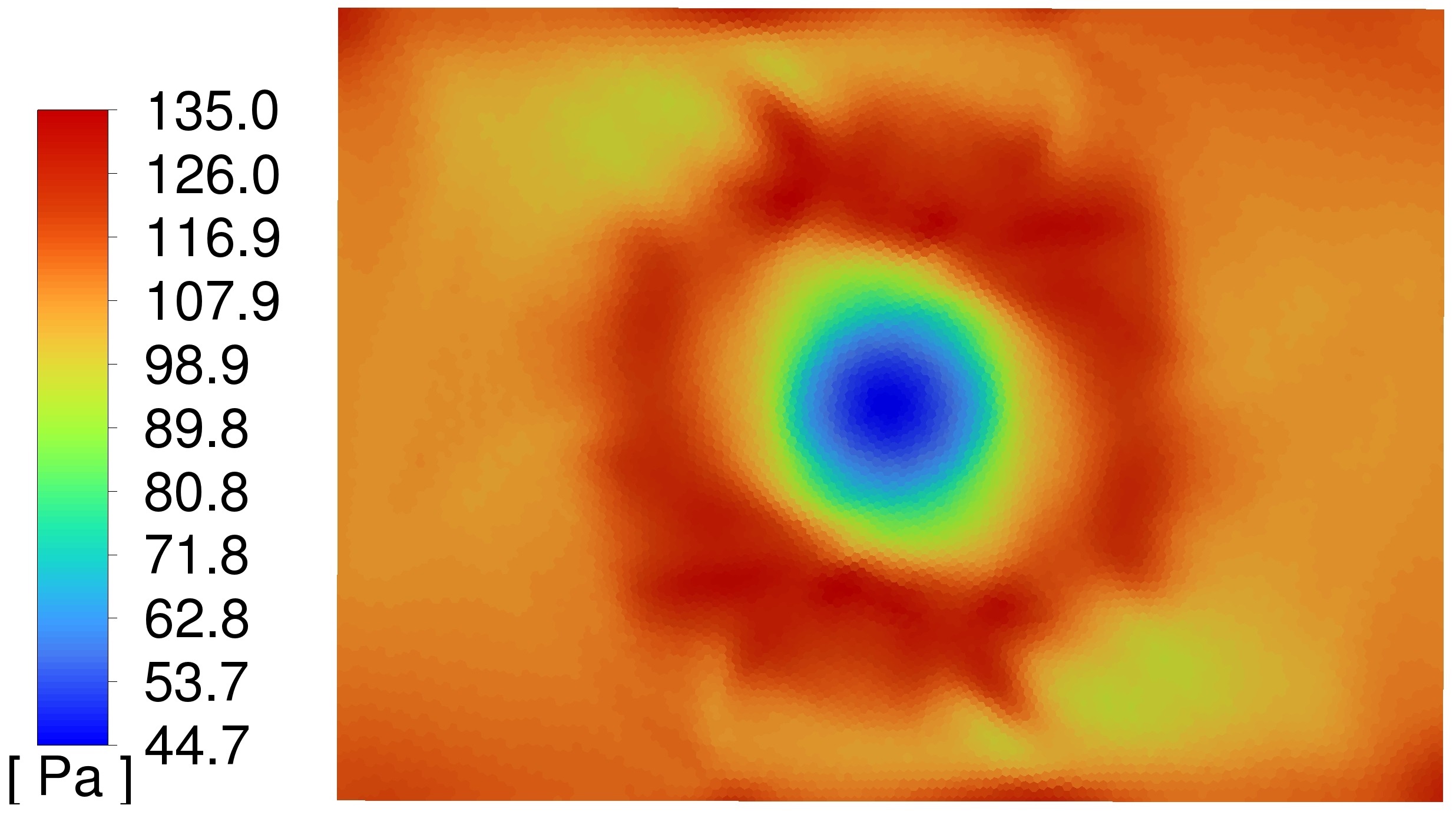}
	\caption{Non-uniform static pressure distribution upstream of tube bundle}
	\label{Figure:NonUniformFlowPressure}
\end{figure}

\begin{figure}[h!]
	\centering
	\includegraphics[width=.95\linewidth]{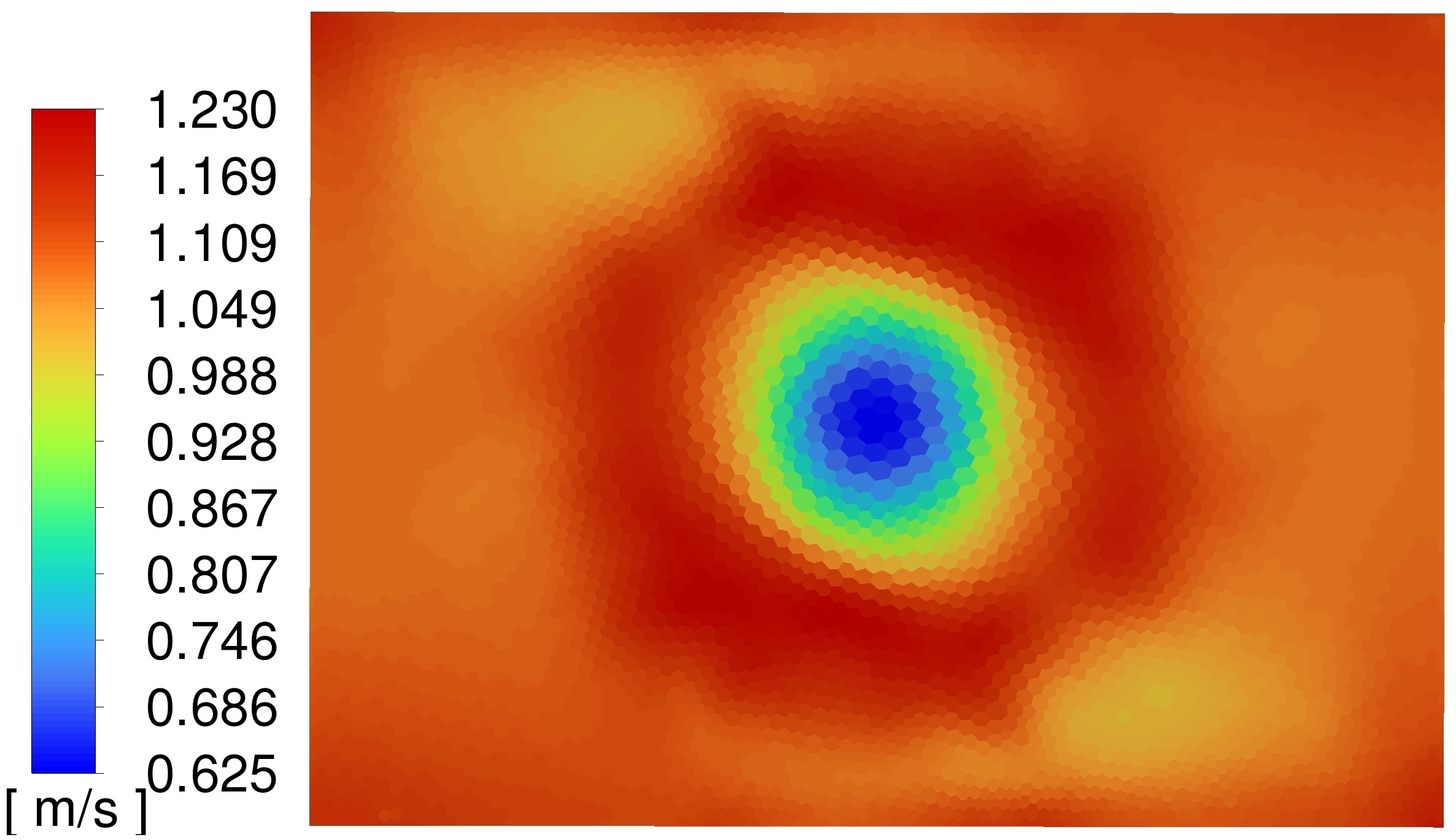}
	\caption{Non-uniform axial velocity distribution downstream of tube bundle}
	\label{Figure:NonUniformFlowVelocity}
\end{figure}

Figure \ref{Figure:NonUniformFlowPressure} illustrates the air pressure distribution obtained just upstream of the tube bundle. Due to this pressure distribution, the cooling air velocity through the finned tube bundle is non-uniform, with Figure \ref{Figure:NonUniformFlowVelocity} showing the velocity of the air leaving the tube bundle. From these results, it is clear that the assumption of uniform cooling air flow is not a good representation of the actual flow field. On the other hand, it is also unclear to what extent the Frozen Rotor method contributes to this non-uniformity, since the position of the rotor relative to the tube bundle is fixed.

Considering the importance of the cooling air uniformity for the current work, other methods of modelling the fan were also tested, discussed in Appendix \ref{Appending:FanModellingApproach}. It was then concluded that the Frozen Rotor approach is the most practical method, while also providing a conservative estimate of the flow uniformity. Therefore, the Frozen Rotor approach was selected for the current work.


\newpage
\section{Results - Effect of Plenum Height}
The effect of a range of plenum chamber heights, from 0.25 to 1 times the fan casing diameter, is demonstrated in this section. This includes the baseline case discussed earlier, with a plenum height of 0.3 times fan casing diameter.

\subsection{Operating Point and Fan Performance}
The operating point flow rate and fan performance is given in Table \ref{Table:PlenumHeight}. Compared to the baseline case, a small improvement in fan performance and cooling air flow rate is possible when a slightly longer plenum chamber is used. At lower or much higher plenum chamber heights, however, the performance is reduced. These trends agree well with that seen in the experimental work by Meyer \& Kr\"{o}ger \cite{Meyer1998}.

\begin{table}[h!]
	\centering
	\caption{Effect of plenum height on performance}
	\label{Table:PlenumHeight}
	\begin{tabular}{c c c c c c}
		& & & & \\
		\hline
		Height & $\dot{V}_F$ (m$^3$/s) & $\Delta p_{ts}$ (Pa) & $P_F$ (kW) & $\eta_{ts}$ (\%) \\
		\hline
		$0.25 \times d_{Fc}$ & 168.8 & 110.5 & 31.99 & 58.32 \\
		$0.3 \times d_{Fc}$ & 169.4 & 110.5 & 31.91 & 58.66  \\
		$0.4 \times d_{Fc}$ & 170.1 & 110.6 & 31.79 & 59.15  \\
		$0.5 \times d_{Fc}$ & 169.9 & 110.2 & 31.76 & 58.94  \\
		$1 \times d_{Fc}$ & 166.7 & 106.8 & 31.76 & 56.05 \\
		\hline
	\end{tabular}	
\end{table}

\subsection{Pressure Distribution}
\begin{figure}[h!]
	\centering
	\begin{tikzpicture}
		\begin{axis}[
			width=\linewidth,
			height=.475\linewidth,
			xlabel=Height (m),
			ylabel=Pressure (Pa),
			xmin=-7,
			xmax=11,
			ymin=-25,
			ymax=170,
			grid=both,
			grid style={draw=gray!15},
			legend style={at={(0.025,0.975)},anchor=north west},
			legend cell align = {left}
			]
			\addplot [color = black,thick,solid] table[col sep=comma,x expr=\thisrowno{0}, y expr=\thisrowno{1}]{pressure_axial_ACHEPorousFine.txt};
			\addplot [color = red,thick,solid] table[col sep=comma,x expr=\thisrowno{0}, y expr=\thisrowno{3}]{pressure_axial_ACHEPorousFine.txt};
			\addplot [color=blue,thick,solid] table[col sep=comma,x expr=\thisrowno{0}, y expr=\thisrowno{2}]{pressure_axial_ACHEPorousFine.txt};
			
			\addplot [color = black,thick,dashed] table[col sep=comma,x expr=\thisrowno{0}, y expr=\thisrowno{1}]{pressure_axial_ACHEPorousExtendedPlenum_1.txt};
			\addplot [color = red,thick,dashed] table[col sep=comma,x expr=\thisrowno{0}, y expr=\thisrowno{3}]{pressure_axial_ACHEPorousExtendedPlenum_1.txt};
			\addplot [color=blue,thick,dashed] table[col sep=comma,x expr=\thisrowno{0}, y expr=\thisrowno{2}]{pressure_axial_ACHEPorousExtendedPlenum_1.txt};
			
			\draw (rel axis cs:0.46,0.975) node[below,align=left] {$0.3 \times d_{Fc}$};
			\draw (rel axis cs:0.74,0.92) node[below,align=left] {$1 \times d_{Fc}$};
			
			\legend{Static, Dynamic, Total}
		\end{axis}
	\end{tikzpicture}
	\caption{Short and long plenum pressure}
	\label{Figure:ExtendedPlenumPressureDistribution}
\end{figure}
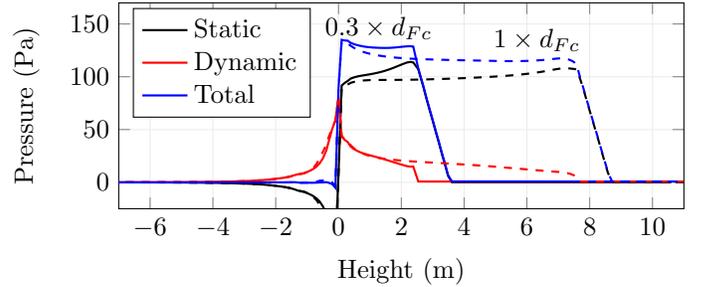

\begin{figure}[h!]
	\centering
	\includegraphics[width=\linewidth]{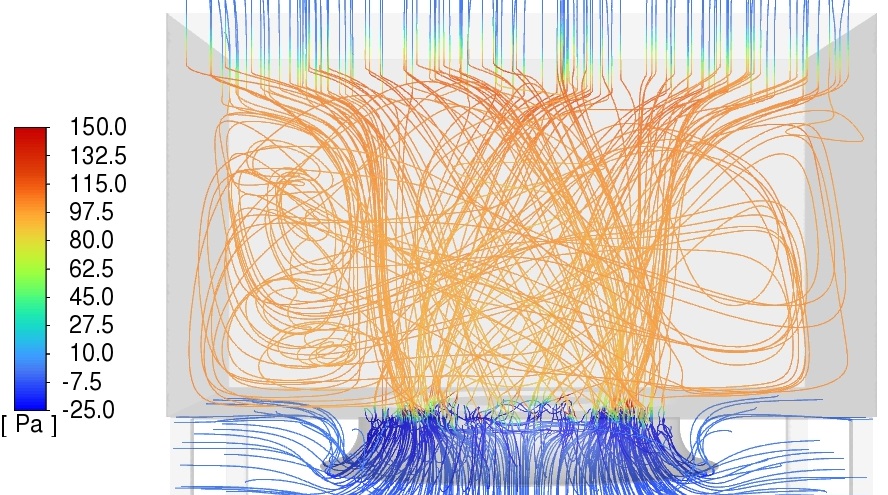}
	\caption{Pathlines colored by static pressure}
	\label{Figure:ACHEPathlinesExtended}
\end{figure}

Figure \ref{Figure:ExtendedPlenumPressureDistribution} illustrates the change in pressure distributions through the ACHE from the baseline case to the longest plenum chamber height tested. Figure \ref{Figure:ACHEPathlinesExtended} illustrates the flow in the extended plenum chamber using pathlines colored by pressure. The distributions are seen to be similar to the baseline case. The total and static pressure is however seen to be lower with the extended plenum, likely due to an increase in duct losses.

\begin{table}[h!]
	\centering
	\caption{Effect of plenum height on axial velocity distribution}
	\label{Table:AxialUniformity}
	\begin{tabular}{c c c c c}
		& & & \\
		\hline
		Height & Min (m/s) & Avg (m/s) & Max (m/s)  \\
		\hline
		$0.25 \times d_{Fc}$ & 0.5099 & 1.094 & 1.243 \\
		$0.3 \times d_{Fc}$ & 0.6253 & 1.096 & 1.228 \\
		$0.4 \times d_{Fc}$ & 0.7502 & 1.098 & 1.229 \\
		$0.5 \times d_{Fc}$ & 0.8216 & 1.095 & 1.240 \\
		$1 \times d_{Fc}$ & 0.8902 & 1.074 & 1.177 \\
		\hline
	\end{tabular}	
\end{table}

\begin{figure}[h!]
	\centering
	\includegraphics[width=.95\linewidth]{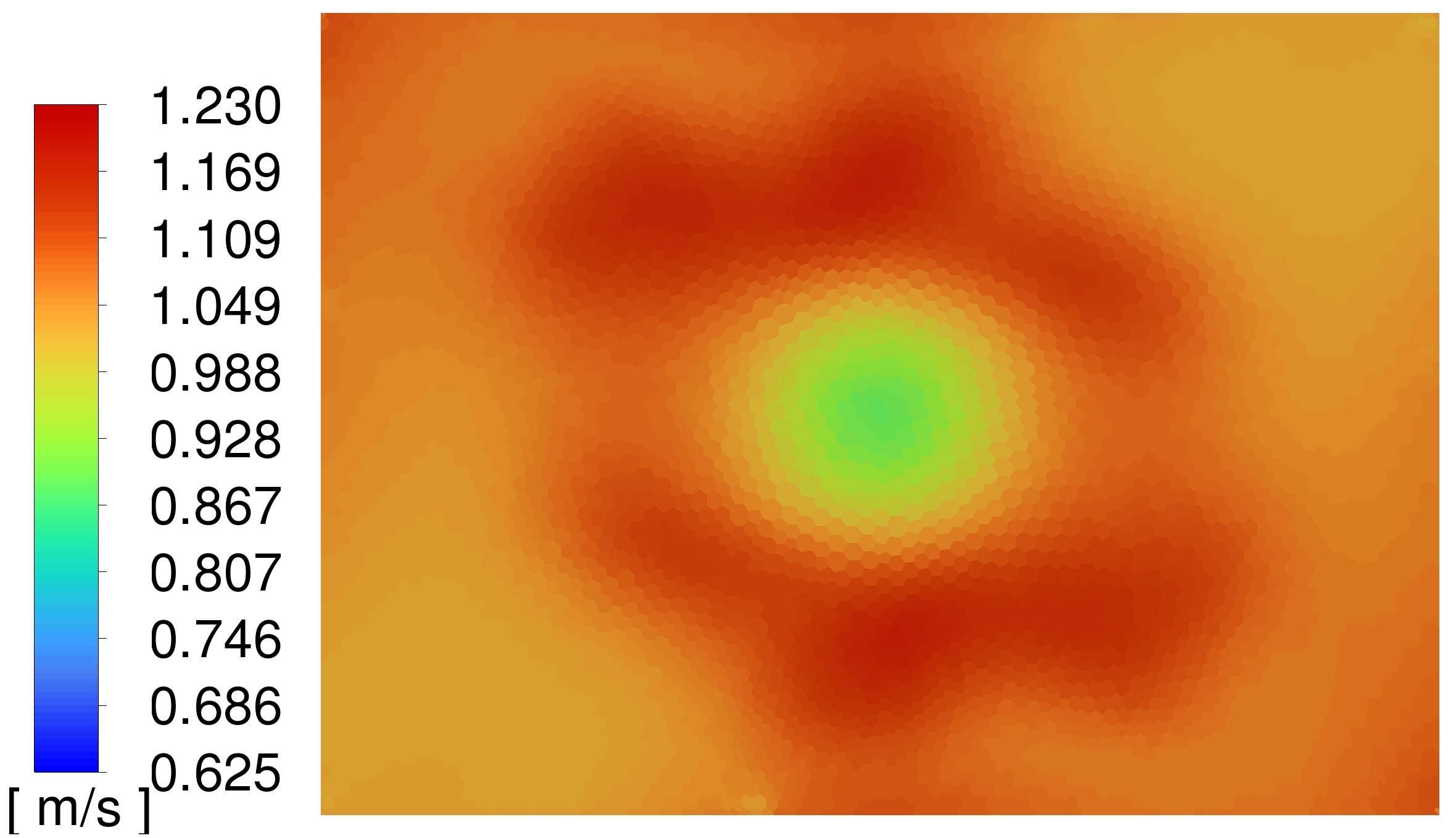}
	\caption{Velocity distribution for long plenum}
	\label{Figure:ExtendedPlenumNonUniformFlowVelocity}
\end{figure}

\subsection{Cooling Air Uniformity}
Table \ref{Table:AxialUniformity} demonstrates the improvement in cooling air flow uniformity through the tube bundle by comparing the average, minimum and maximum axial velocities at the tube bundle outlet. Figure \ref{Figure:ExtendedPlenumNonUniformFlowVelocity} illustrates the axial velocity distribution downstream of the tube bundle resulting from the longest plenum chamber. When compared to Figure \ref{Figure:NonUniformFlowVelocity} for the baseline case, it is clear that a significant improvement in flow uniformity has been achieved.

\section{Conclusions and Recommendations}
This paper documents the devolopment of a CFD model to simulate the air-side interaction between an axial flow cooling fan and an air-cooled heat exchanger for an sCO$_2$ Brayton cycle CSP plant. Due to the sensitivity of sCO$_2$ to changes in temperature, the focus is not only on accurately modelling the operating point flow rate obtained, but also on the prediction of the uniformity of the cooling air. 

Good agreement is seen between the baseline case CFD results and the analytical prediction, with a difference of only 0.5341\% between the analytical and simulated flow rates. Good agreement is also found between the fan performance obtained at this operating point and the Fan Test Facility CFD results from previous work \cite{Boshoff2023}. This indicates that the CFD model accurately predicts the operating point and fan performance. 

A comparison of various fan modelling methods also indicates that the Frozen Rotor approach gives a good indication of the uniformity of the flow. While the non-uniformity is slightly exagerated due to the position of the fan rotor being fixed, it is considered to be a conservative and therefore safe prediction. The Frozen Rotor approach can also be completed within a relatively short time frame when compared with a transient method. Based on these findings, the model is considered to be a practical and accurate method of simulating ACHEs for sCO$_2$ power cycles.

Furthermore, for this particular case, the results from the plenum chamber height simulations indicate that a plenum chamber height between 0.3$\times$ and 0.5$\times$ the fan casing diameter provides the highest cooling air flow rate. On the other hand, a longer plenum length is shown to improve the uniformity of the flow significantly, particularly behind the large hub. A plenum length of about 0.5$\times$ the fan casing diameter therefore seems to be a good combination of both cooling rate and uniformity.

Unfortunately it is still unclear to what extent the non-uniform flow will affect the sCO$_2$-side flow and the thermal performance. It is therefore recommended that the current work be continued, with the focus being on the modelling of the sCO$_2$-side behaviour and the heat transferred between the fluids. Such an investigation could determine if over- or undercooling of the sCO$_2$ occurs in the tube bundle, and would give an indication of the stability and thermal performance of the heat exchanger.

\section{Acknowledgements}
The authors would like to acknowledge the following groups and individuals for their contribution towards the completion of this work: The Solar Thermal Energy Research Group (STERG) at Stellenbosch University for their ongoing support, Professor Alessandro Corsini, Dr Giovanni Delibra and Dr Lorenzo Tieghi from Sapienza University of Rome, for their guidance during the early stages of the work, and the South African Department of Science and Innovation's CHPC, for providing the computational resources required to complete the simulations.

\bibliographystyle{asmems4}
\bibliography{ASME TE 2024}

\appendix
\vspace{20 pt}
\noindent
\textbf{\LARGE Appendices}
\section{Sensitivity Analysis}
\label{Appendix:SensitivityAnalysis}

\subsection{Mesh Resolution}
The sensitivity to mesh density was tested by succesively refining the mesh. The fan performance obtained for each case is given in Table \ref{Table:PorousZoneMeshSensitivity}, with small changes seen between the final two refinements. Therefore, the element size at the final refinement is used for all simulations.

\subsection{Inlet Height}
The sensitivity to the inlet height was tested by both decreasing and increasing the inlet bellmouth height. The results are given in Table \ref{Table:PorousZoneInletSensitivity}, which demonstrates almost no change within this range. An inlet height of 1$\times d_{Fc}$ is therefore used for all simulations.

\subsection{Outlet Height and Boundary Condition}
The sensitivity to the outlet sub-domain size was tested by increasing the height of the outlet sub-domain. Furthermore, since the vertical side surfaces of the outlet sub-domain are specified as pressure outlet boundary conditions, the pressure at the sides directly downstream of the porous zone are fixed at zero. It was unclear if this proximity could affect the pressure distribution significantly. Therefore, a second outlet sub-domain variant splits the outlet surface into two parts: the vertical side surfaces are specified as no-shear wall surfaces, which allows the pressure directly downstream of the bundle to vary, while the top horizontal surface is kept as a pressure outlet. The results in Table \ref{Table:PorousZoneOutletSensitivity} shows almost no change between these variants. The first outlet sub-domain variant with a height of 1$\times d_{Fc}$ is therefore used for all simulations.
\begin{table}[h!]
	\centering
	\caption{Mesh sensitivity}
	\label{Table:PorousZoneMeshSensitivity}
	\begin{tabular}{c c c c c c}
		& & & & \\
		\hline
		Elements & $\dot{V}_F$ (m$^3$/s) & $\Delta p_{ts}$ (Pa) & $P_F$ (kW) & $\eta_{ts}$ (\%) \\
		\hline
		4533042 & 168.3 & 109.8 & 32.71 & 56.47 \\
		5597236 & 166.2 & 107.9 & 32.15 & 55.77 \\
		7501652 & 170.0 & 111.0 & 32.01 & 58.93 \\
		8510751 & 169.4 & 110.5 & 31.91 & 58.66 \\
		\hline
	\end{tabular}	
\end{table}
\begin{table}[h!]
	\centering
	\caption{Sensitivity to inlet height}
	\label{Table:PorousZoneInletSensitivity}
	\begin{tabular}{c c c c c c}
		& & & & \\
		\hline
		Height & $\dot{V}_F$ (m$^3$/s) & $\Delta p_{ts}$ (Pa) & $P_F$ (kW) & $\eta_{ts}$ (\%) \\
		\hline
		$0.75 \times d_{Fc}$ & 169.4 & 110.4 & 31.91 & 58.63 \\
		$1 \times d_{Fc}$ & 169.4 & 110.5 & 31.91 & 58.66 \\
		$2 \times d_{Fc}$ & 169.5 & 110.5 & 31.91 & 58.70 \\
		\hline
	\end{tabular}	
\end{table}
\begin{table}[h!]
	\centering
	\caption{Sensitivity to outlet height}
	\label{Table:PorousZoneOutletSensitivity}
	\begin{tabular}{c c c c c c c}
		& & & & & \\
		\hline
		Height & Variant & $\dot{V}_F$ (m$^3$/s) & $\Delta p_{ts}$ (Pa) & $P_F$ (kW) & $\eta_{ts}$ (\%) \\
		\hline
		$1 \times d_{Fc}$ & 1 & 169.4 & 110.5 & 31.91 & 58.66 \\
		$2 \times d_{Fc}$ & 1 & 169.4 & 110.5 & 31.90 & 58.68 \\
		$1 \times d_{Fc}$ & 2 & 169.4 & 110.5 & 31.90 & 58.68 \\
		\hline
	\end{tabular}	
\end{table}

\newpage
\section{Comparison of Fan Modelling Methods}
\label{Appending:FanModellingApproach}
The Frozen Rotor approach used in the current work seems to accurately predict the cooling air flow rate and fan performance. However, it is a steady approximation of the actual flow field, which fixes the fan blade position relative to the tube bundle. It exaggerates the effect of blade pressure pulses, and therefore predicts a highly non-uniform velocity profile through the tube bundle. To ensure that this method is appropriate, it is compared with two variations of fan modelling approaches, using a simplified no tip gap case.

\subsection{Frozen Rotor with Tangential Offset}
The sensitivity of the results to a change in blade position is tested firstly, using the baseline case with the fan rotated in the tangential direction by 22.5°.

\subsection{Mixing Plane}
The Mixing Plane approach is another steady flow approximation, which works by averaging the flow field in the tangential direction at one boundary, after which the boundary condition on the other side is updated. To use this method, the mesh interface surfaces connecting the fan rotor and the plenum chamber sub-domains are replaced with a pressure outlet boundary condition on the upstream side, and a mass flow inlet boundary condition on the downstream side. The mixing plane is then set up using these two surfaces. This method therefore provides a more uniform cooling air distribution through the tube bundle.



\subsection{Results}
\begin{figure}[h!]
	\centering
	\begin{minipage}{0.49\linewidth}
		\centering
		\includegraphics[width=.95\linewidth]{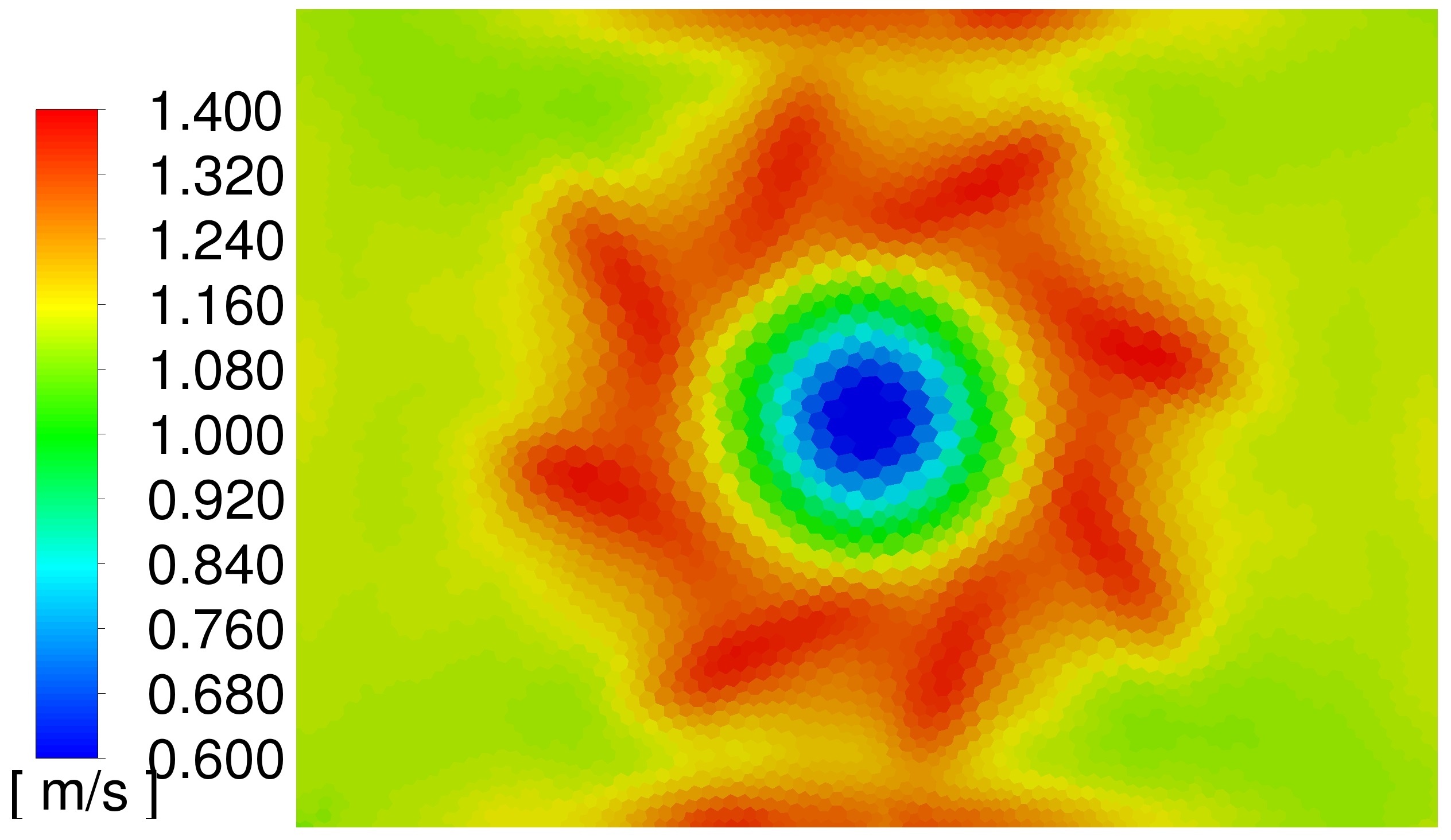}
	\end{minipage}
	\begin{minipage}{0.49\linewidth}
		\centering
		\includegraphics[width=.95\linewidth]{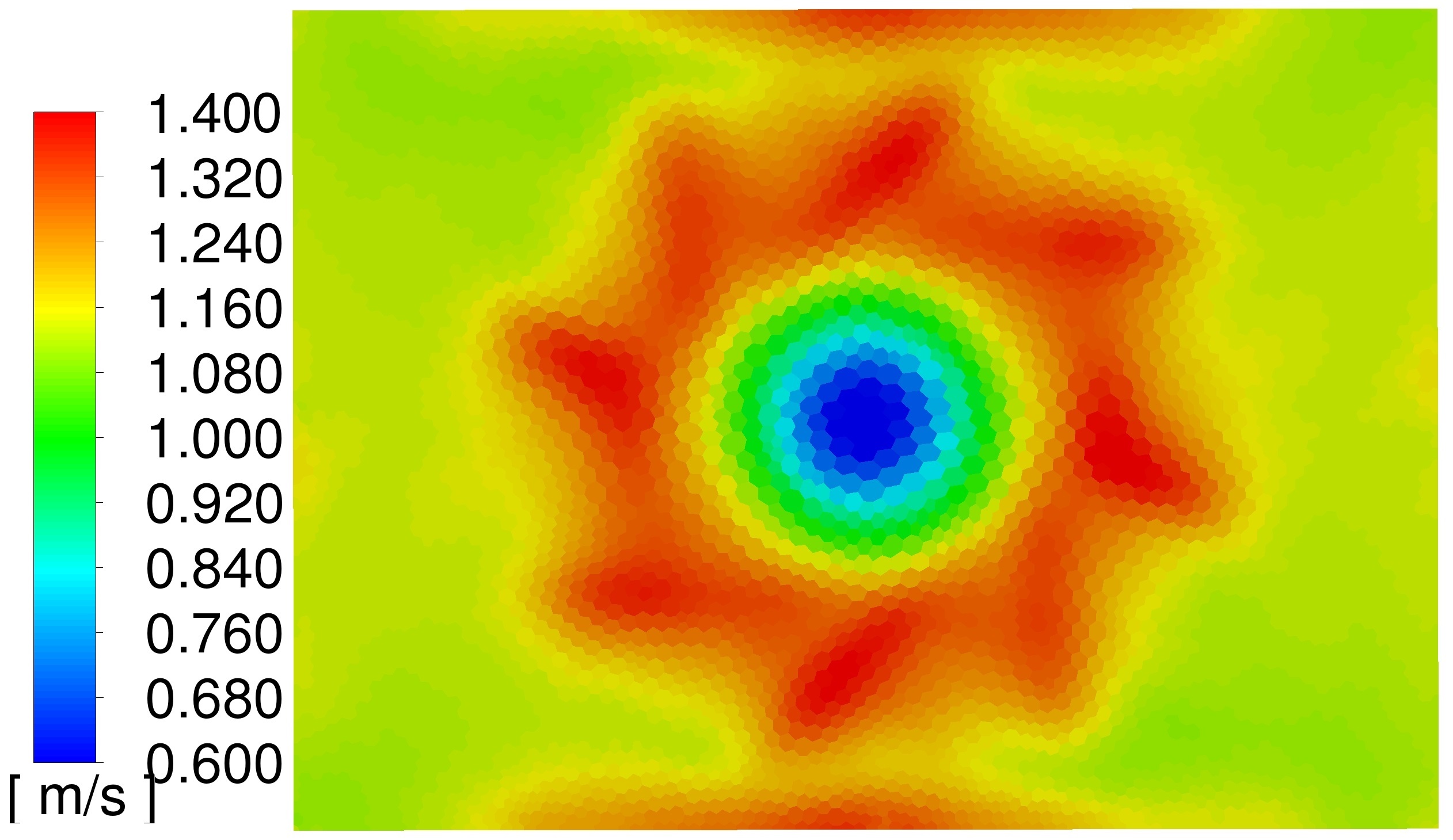}
	\end{minipage}
	\caption{Frozen rotor approach with no tangential offset (left) and a 22.5° offset (right)}
	\label{Figure:NonUniformFlowVelocityFrozen}
\end{figure}

\begin{figure}[h!]
	\centering
	\includegraphics[width=.4655\linewidth]{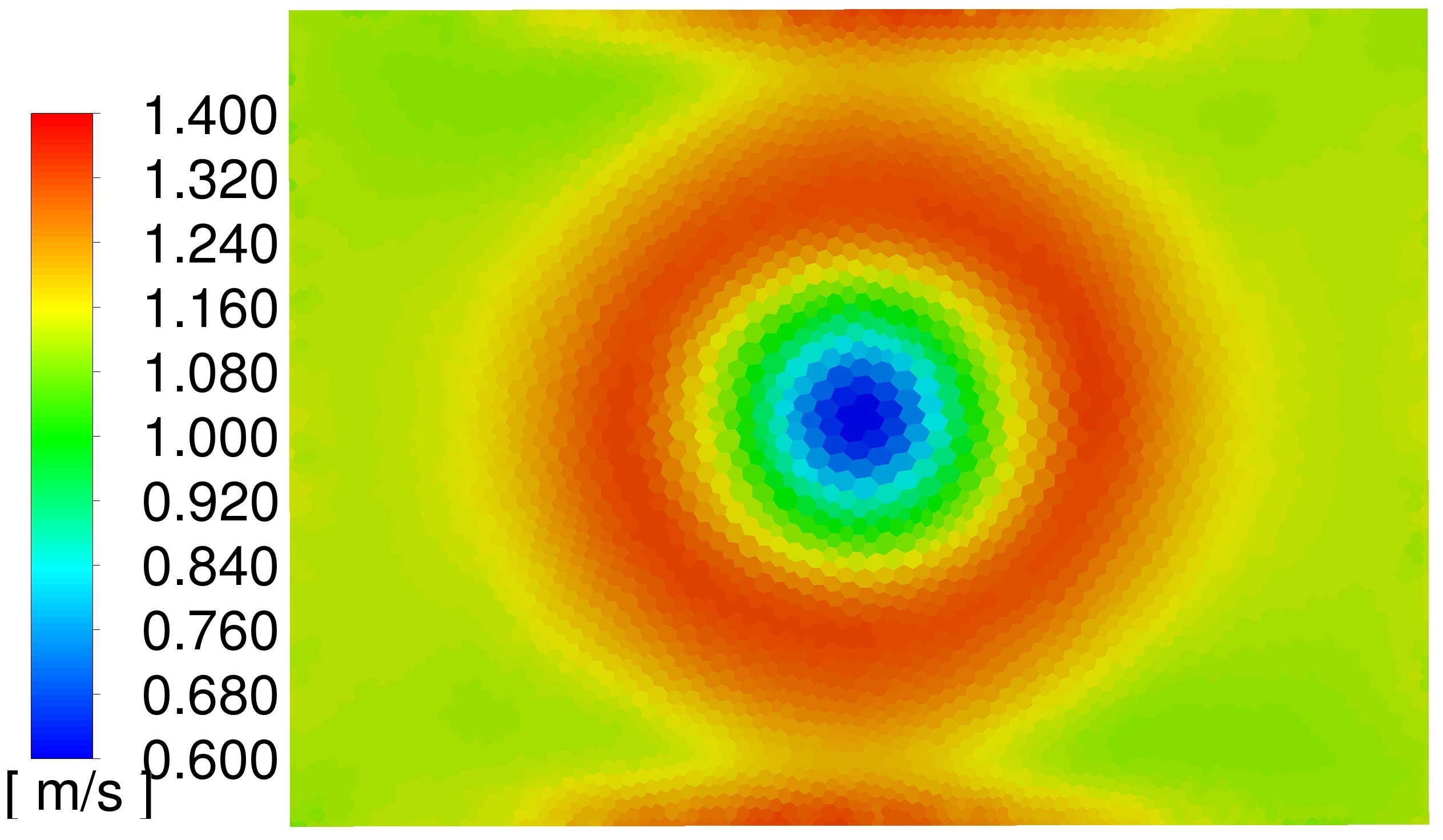}
	\caption{Mixing plane approach}
	\label{Figure:NonUniformFlowVelocityMixingSliding}
\end{figure}

The resulting fan performances given in Table \ref{Table:FanModellingMethodsPerformance} are seen to be similar. Figures \ref{Figure:NonUniformFlowVelocityFrozen} and \ref{Figure:NonUniformFlowVelocityMixingSliding}, however, demonstrate the differences between cooling air distributions predicted by each method. It is clear that the fan modelling method used significantly influences the flow uniformity. Since the frozen rotor method is considered to provide a conservative estimate of the non-uniformity of the flow, it was selected for the current work.

\begin{table}[h!]
	\centering
	\caption{Fan performance for no tip gap case}
	\label{Table:FanModellingMethodsPerformance}
	\begin{tabular}{c c c c c c}
		& & & & \\
		\hline
		Method & $\dot{V}_F$ (m$^3$/s) & $\Delta p_{ts}$ (Pa) & $P_F$ (kW) & $\eta_{ts}$ (\%) \\
		\hline
		FR 0° & 181.4 & 123.9 & 35.63 & 63.08 \\
		FR 22.5° & 181.6 & 123.9 & 35.62 & 63.17 \\
		MP & 181.6 & 122.9 & 35.55 & 62.80 \\
		\hline
	\end{tabular}	
\end{table}




\newpage

\end{document}